\documentclass[sigconf]{acmart}
\usepackage{booktabs} 
\usepackage{graphicx}
\usepackage{fancybox}
\usepackage{balance}
\usepackage{multirow}
\definecolor{LightYellow}{HTML}{FFFFC7}
\colorlet{LightRed}{red!50!}
\newcolumntype{g}{>{\columncolor{white}}c}
\let\oldbibitem\bibitem
\def\bibitem{\vfill\oldbibitem}

\newcommand{\hypobox}[1]{

        \begin{center}\noindent\thicklines\setlength{\fboxsep}{8pt}\cornersize{0.2}\ovalbox{

                \begin{minipage}{3.0in}

                        \textit{#1}

                \end{minipage}} 

        \end{center}} 

\newcommand{\datadoi}{%
  \begingroup\normalfont
  \smash{\href{https://doi.org/10.5281/zenodo.3858046}{\includegraphics[height=1.3\fontcharht\font`\B,trim=0 3 0 0]{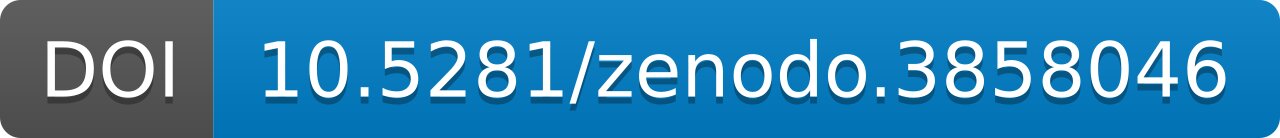}}}%
  \endgroup
}   

\copyrightyear{2020}
\acmYear{2020}
\setcopyright{acmlicensed}\acmConference[ESEM '20]{ESEM '20: ACM / IEEE International Symposium on Empirical Software Engineering and Measurement (ESEM)}{October 8--9, 2020}{Bari, Italy}
\acmBooktitle{ESEM '20: ACM / IEEE International Symposium on Empirical Software Engineering and Measurement (ESEM) (ESEM '20), October 8--9, 2020, Bari, Italy}
\acmPrice{15.00}
\acmDOI{10.1145/3382494.3410685}
\acmISBN{978-1-4503-7580-1/20/10}

\begin{document}
\title{Effect of Technical and Social Factors on Pull Request Quality for the NPM Ecosystem}

\author{Tapajit Dey}
\email{tdey2@vols.utk.edu}
\affiliation{%
  \institution{The University of Tennessee}
  \streetaddress{1520 Middle Dr.}
  \city{Knoxville}
  \state{TN}
  \country{USA}
  \postcode{37996}
}
\author{Audris Mockus}
\email{audris@utk.edu}
\affiliation{%
  \institution{The University of Tennessee}
  \streetaddress{1520 Middle Dr.}
  \city{Knoxville}
  \state{TN}
  \country{USA}
  \postcode{37996}
}


\begin{abstract}
Pull request (PR) based development, which is a norm for the social coding platforms, entails the challenge of evaluating the contributions of, often unfamiliar, developers from across the open source ecosystem and, conversely, submitting a contribution to a project with unfamiliar maintainers. Previous studies suggest that the decision of accepting or rejecting a PR may be influenced by a diverging set of technical and social factors, but often focus on relatively few projects, do not consider ecosystem-wide measures, or the possible non-monotonic relationships between the predictors and PR acceptance probability. 
We aim to shed light on this important decision making process by testing which measures significantly affect the probability of PR acceptance on a significant fraction of a large ecosystem, rank them by their relative importance in predicting PR acceptance, and determine the shape of the functions that map each predictor to PR acceptance. 
We proposed seven hypotheses regarding which technical and social factors might affect PR acceptance and created 17 measures based on them. Our dataset consisted of 470,925 PRs from 3349 popular NPM packages and 79,128 GitHub users who created those. We tested which of the measures affect PR acceptance and ranked the significant measures by their importance in a predictive model.
Our predictive model had and AUC of 0.94, and 15 of the 17 measures were found to matter, including five novel ecosystem-wide measures. Measures describing the number of PRs submitted to a repository and what fraction of those get accepted, and signals about the PR review phase were most significant.
We also discovered that only four predictors have a linear influence on the PR acceptance probability while others showed a more complicated response.
\end{abstract}

\ccsdesc[300]{Software and its engineering~Open source model}
\ccsdesc[500]{Human-centered computing~Open source software}
\ccsdesc[500]{Human-centered computing~Empirical studies in collaborative and social computing}

\keywords{Pull Request, NPM Packages, Predictive Model, Social Factors}

\maketitle

\section{Introduction}\label{s:intro}
With the advent of social-coding platforms like GitHub~\cite{dabbish2012social},
the Pull Request (PR) based development model has become the norm. This model allows
developers outside of a project to contribute without compromising the quality of the
original project by only merging approved changes to the repository, and was found to
be associated with shorter review times and larger numbers of contributors compared
to mailing list code contribution models~\cite{zhu2016effectiveness}. 

However, the PR based development model, while clarifying and simplifying the process of contributions from unfamiliar outsiders, does not, by itself, ensure the  quality of contributions.  
This entails the challenges of evaluating the 
quality of the contributed code (a task that is typically performed by a PR
integrator) and maintaining the quality of the project~\cite{gousios2015work}. The PR
integrators of popular projects are often overloaded with reviewing multiple pull
requests~\cite{gousios2015work}, which adds additional complexity to the process.
It has been extensively  documented (see, e.g.~\cite{xie2013impact}) that large numbers 
of low-quality issues may overwhelm the projects and the same is true for PRs. 
Being able to gauge the quality of a submitted PR may, therefore, benefit the
integrators to prioritize their review process and may, consequently, increase the 
efficiency of the process. However, PR ``quality'' has no universal definition and may mean different things in different contexts. We, therefore, chose
the ultimate pragmatic indicator: whether it is merged (accepted)
or not. It should be based on the contextual knowledge of
the integrator at the time of acceptance and should take into
account a variety of factors the integrator has to consider when
accepting a PR. By doing this we follow a comprehensive treatment of code contribution
theory in Rigby et. al.~\cite{rigby2014peer} that considers the acceptance
rate as one of the most fundamental properties of the peer-review
systems. In this paper we, therefore, define the quality of a PR by
its probability of getting merged.

Several studies (e.g. ~\cite{weissgerber2008small,soares2015acceptance,rahman2014insight,jiang2013will,mockus2002two,baysal2012secret},) 
investigated the effects of various technical and social factors on the 
PR acceptance probability. However, those studies primarily 
contained relatively few ($<$ 100) projects, potentially limiting their generalizability. 
Different studies highlighted different factors that significantly influence the PR acceptance probability with no clear answer as to what factors apply broadly. Most studies did not present the relative importance of these factors, nor did they report the functional relationship between the PR acceptance probability and values of the predictors. Finally, prior studies that focused on a set of specific projects did not take into account the ecosystem-wide nature of developer participation (and the corresponding experience).

To address these gaps, 
we analyzed how different technical and social factors related to the characteristics of a PR, the 
PR creator, the repository to which the PR is submitted, the social proximity  between 
the PR creator and the repository, and the PR review phase influence its probability of
being accepted by analyzing 470,925 PRs for 3349 packages (2740 different GitHub  repositories), each with
more than 10,000 monthly downloads and at least 5 PRs created against their
GitHub repository, from the NPM ecosystem, which is one of the largest open source
software ecosystems at present. 

Our goal in this study is to deepen the understanding of how various social and technical
factors influence the PR acceptance probability at an ecosystem level (where we characterize projects, developers, and PRs based on measures obtained not just for the specific set of projects, but on the entire ecosystem of projects and developers in the ecosystem-wide software supply chain~\cite{amreen2019methodology}). Knowing how these measures, including the contextual factors that may be unique to individual projects or sub-ecosystems, influence PR acceptance may 
help the PR creators, integrators, and also the tool designers who design pull
request evaluation interfaces.  Specifically, the tool designers might choose to make important signals
more readily available to the PR creators, who can better format their individual pull
requests to have a better chance of having their contributions accepted, and to the PR integrators,
who can look for those signals to gauge the quality of the PRs they are evaluating. \footnote{We are not performing a causal analysis, thus, we do not intend
to make the integrators believe that a PR is bad just because it doesn't match the
characteristics of a typically acceptable PR, or vice versa.} 

Our key contributions include the conceptual replication of prior findings for the 
ecosystem-level model and data that relates the probability of acceptance to the technical characteristics of a PR, the track record of a PR creator in terms of having their PRs accepted and their social proximity  to the repository to which the PR is submitted, and measures describing the characteristics of the PR review phase. 
We also introduce new ecosystem-level measures: \textit{the overall experience of the PR creator across all OSS projects, the leniency of a repository in terms of accepting PRs,} and \textit{the presence of a dependency between the repository the PR creator submitted the PR to  and the projects they previously contributed to} and show that they have a significant impact. A predictive model with these measures achieved an AUC value of 0.94. Finally, we observed nonlinear dependencies between the PR acceptance probability and most of these measures (see Section~\ref{ss:partial}), suggesting a variety of potential mechanisms and other factors like the presence of bots in the dataset that appear to drive PR acceptance in different contexts.
We have also created a dataset with the curated data of the PR properties we measured, along with their descriptions, a code snippet for creating a Random Forest model using the data, and we also included the final Random Forest model we used for predicting PR acceptance. The dataset is available at: \datadoi~\cite{tapajit_dey_2020_3858046}. 

The rest of the paper is structured as follows: We discuss the related works about this
topic in Section~\ref{s:relwork} and present the specific research questions we address 
in the study in Section~\ref{s:rq}. The data along with the methodology for data
collection, processing, and analysis is discussed in Section~\ref{s:method}. We present 
and discuss the results of our study in Section~\ref{s:result}. The limitations to our 
study and future works are discussed in Section~\ref{s:limit}, and we conclude our paper 
in Section~\ref{s:conclusion}.

\section{Related Works}\label{s:relwork}
There have been a good number of studies on PRs, which investigated different
questions related to the dynamics of PR creation and acceptance, e.g. estimation of PR
completion times~\cite{maddila2019predicting}, finding the right evaluator for a 
particular PR~\cite{yu2014should,yu2014reviewer, jiang2017should,yu2016reviewer},
predicting whether a PR will see any activity within a given time window
~\cite{prioritizer}, how PRs get rejected~\cite{steinmacher2018almost} etc. There are other studies that explore the perspective of the PR
creators~\cite{gousios2016work} and the PR integrators~\cite{gousios2015work}, and list 
the challenges and practices in PR creation and merging scenarios. 

A number of studies describe various factors that influence the chance of a PR getting accepted, like~\cite{soares2015acceptance}, which advocates using association rules to find the important factors, and found that the acceptance rates vary with the language the repository is written in, and also that having fewer commits, no additions, some deletions, some changed files, and the author having created a PR before and/or being part of the core team increase the chance of getting a PR accepted;~\cite{weissgerber2008small}, which indicates smaller PRs are more likely to get accepted;~\cite{yu2015wait}, which shows previously established track records of the contributors, availability and workload of the evaluators, and continuous integration based automated testing etc. have an impact on the latency of PR evaluation;~\cite{rahman2014insight}, which examined the effects of developer experience, language, calendar time etc. on the PR acceptance;~\cite{dey2020representation}, which showed that the specific expertise of developers might influence PR acceptance probability;~\cite{tsay2014influence}, which analyzed   the association of various technical and social measures, e.g. adherence to contribution norms, social proximity  between the creator and the project, amount of discussion around the PR, number of followers of the PR creator, and popularity of the repository to which the PR is submitted,  with  the  likelihood  of  PR  acceptance.
There are a number of case studies that discuss the PR acceptance scenario in various OSS projects, like the Linux kernel~\cite{jiang2013will}, Firefox~\cite{baysal2012secret,mockus2002two}, Apache~\cite{mockus2002two} etc. 
All of the studies mentioned above, except~\cite{tsay2014influence},  focused on a few
($<$100) projects, so the general applicability of their findings wasn't verified at an
ecosystem level. As for~\cite{tsay2014influence}, while it studied a large number of
(12,482) GitHub projects and reported the significance of various social and technical
factors by the \textit{odds ratio} measure (which itself has a few drawbacks, e.g. odds ratio can overstate the effect size~\cite{davies1998can} and can lead to misleading implications, especially when events, e.g. PR acceptance in this situation, are very common or very  rare~\cite{davies1998can,altman1998odds}), it doesn't show the relative importance of the factors involved, nor does it show how the PR acceptance probability varies with the values of these factors. 

In our study, we aim to add to the current body of knowledge about this topic by addressing the limitations of the previous studies, specifically by showing the relative importance of different factors, identifying how different values of each of these factors affect the PR acceptance probability, and verify the applicability of our findings for the NPM ecosystem.

\section{Research Questions}\label{s:rq}

Our first research question focuses on identifying various social and technical factors that might affect the PR acceptance probability, and testing the significance of those factors for our dataset:\\
\textbf{RQ1: What are the various Social and Technical factors that affect PR acceptance probability?}\\
We formulated several hypotheses for addressing this research question. In particular, our review of the related literature revealed a number of factors that might potentially affect PR acceptance probability, viz.:
\begin{enumerate}
    \item [H1] \textit{The technical characteristics of a PR}, described by its size and factors like inclusion of test code and issue fixes, \textit{have a significant impact on PR acceptance probability,} with existing literature (e.g. ~\cite{weissgerber2008small,tsay2014influence}) suggesting that smaller PRs are more likely to be accepted as they are easier to review and more likely to involve a single task.
    \item [H2] \textit{Social proximity between the PR creator and the repository to which it is created also increases the PR acceptance probability} (see, e.g.~\cite{tsay2014influence,yu2015wait}).
    \item [H3] \textit{A previous track record of the PR creator in getting their contributions accepted (in the same ecosystem) increases the PR acceptance probability} (see, e.g.~\cite{prioritizer}).
    \item [H4] \textit{Characteristics of the PR review phase can have an impact on the probability of PR acceptance}, e.g. a higher amount of discussion around the PR was found to have a negative effect on PR acceptance~\cite{tsay2014influence}.
\end{enumerate}
In addition to considering the particular measures related to these factors described in earlier studies, we hypothesized that a few other measures might also affect the PR acceptance probability:
\begin{itemize}
    \item [H5] \textit{A more experienced PR creator will have a higher chance of getting their PRs accepted}, because experienced developers would create better PRs, e.g, via H1. The experience may also be associated with their reputation, thus potentially reducing the social distance via H2. It is worth mentioning that~\cite{rahman2014insight} also described a variable named developer experience that can affect the PR acceptance probability, however, while they simply referred to the duration of their job experience, we refer to a set of more specific measures, viz. the number of commits made by them and the number of projects they have contributed to in the entire OSS ecosystem.
    \item [H6] \textit{PRs to a repository, which has a track record of being more lenient in terms of accepting PRs, are more likely to be accepted}, as evidenced by the fact that getting a PR accepted in projects like Linux kernel tends to be harder due to the project practices of placing stringent requirements on the PRs.
    \item [H7] \textit{If any of the projects the PR creator previously contributed to  depend on the repository to which the PR is being created, it is more likely to be accepted}, since it may be in the self-interest of the PR creator to make the quality of the patch better if their projects depend on the repository. Alternatively, depending on a specific repository may increase their expertise on that project. 
\end{itemize}
\textit{Our first research question is addressed by testing the validity of the above-mentioned hypotheses for our dataset.} Specifically, testing H1-H4 represents a conceptual replication of earlier work on a different dataset with ecosystem-wide operationalizations of the original measures.

We also want to investigate the relative importance of the factors mentioned above in predicting if a PR will be accepted, since such knowledge would help the developers to know and prioritize the aspects they should focus on to get their contributions accepted (for the PR creators), or to gauge the quality of a submitted PR (for the integrators), or trying to decide which signals to make available to the parties involved (for tool designers).\\
\textbf{RQ2: What are the relative importance of different measures related to the factors found to be significant while trying to predict whether a PR will be accepted?}

Finally, we want to find out the functional relationship between the PR acceptance probability and the aforementioned predictors. Should developers try to maximize or minimize them, how important such modification might be in increasing the chances, or is there a ``sweet spot'' to target? This can also help the researchers in gaining a better understanding of the complex dynamics of the process of PR creation and acceptance.\\
\textbf{RQ3: How does the PR acceptance probability vary with different values of the predictors?}

\section{Methodology}\label{s:method}
In this section we describe the data used in this study, the process of data collection and preprocessing, and the measures used for describing the factors mentioned above.

\subsection{PR Selection}
To conduct an empirical study investigating how different technical and social factors affect the PR quality we chose to focus on the node package manager (NPM) ecosystem because of its size, popularity among software developers at present, and the availability of data. NPM is a package manager of JavaScript packages, and is one of the largest OSS communities at present, with over a million different packages and millions of users (estimated 4 million in 2016~\cite{npmuser}, and about 4000 new users on an average day\footnote{https://twitter.com/seldo/status/880271676675547136}, and is a focus of a number of studies that investigated several aspects of the ecosystem, e.g. the dependency networks in NPM~\cite{decan2018impact}, NPM package popularity~\cite{dey2018software}, issues raised against the packages~\cite{dey2018modeling,dey2019patterns,dey2020deriving}, problems associated with library migration~\cite{zapata2018towards} etc.
However, most packages in NPM are not widely used and have limited or no PRs. We, therefore, focused on the NPM packages with over 10,000 monthly downloads (the ``popular'' packages) since January, 2018 (to ensure that they maintained their popularity for a sustained period of time), that also has an active GitHub repository with at least 5  PRs created against it. We chose to remove packages with very few PRs because the effects of the repository related measures on PR acceptance might give misleading results for them. In addition, we chose to only focus on the PRs that were created on or before April, 2019 and were marked as ``closed'', to ensure that they have had sufficient time to be resolved. This way, we were left with 470,925 PRs that were created against 2740 GitHub repositories of 3349 NPM packages. These PRs were created by 79,128 unique GitHub users, and were evaluated by 3633 unique integrators.

\begin{table*}[ht]
\caption{Detailed Definition of the selected Measures and their Descriptive Statistics: Median, Mean, 5\% and 95\% values for continuous variables, number of Yes/No values (represented by 0 \& 1 in the data, respectively) for binary variables (highlighted in yellow). Measures marked with asterisk(*) were not used in any previous study.}
\label{t:vars}
\resizebox{\linewidth}{!}{%
\begin{tabular}{p{4.5cm}lp{6cm}llll}
\toprule
\textbf{Underlying Factors (Hypothesis)} & \textbf{Measure} & \textbf{Variable Description} & \textbf{5\%} & \textbf{Median} & \textbf{Mean} & \textbf{95\%} \\ \midrule
\multirow{6}{4cm}{PR Characteristics (H1)} & additions & Number of lines added in the PR & 1 & 12 & 703 & 619 \\
 & deletions & Number of lines deleted in the PR & 0 & 2 & 385 & 248 \\
 & commits & Number of commits in the PR & 1 & 1 & 4 & 7 \\
 & changed\_files & Number of files modified in the PR & 1 & 2 & 10 & 17 \\
 \rowcolor{LightYellow}
 \cellcolor{white} & contain\_issue\_fix & If the PR contained a fix for an issue & \multicolumn{4}{l}{\textbf{No (0)}: 267811 (57\%), \textbf{Yes (1)}: 203114 (43\%)} \\
 \rowcolor{LightYellow}
 \cellcolor{white} & contain\_test\_code & If the PR contained test code & \multicolumn{4}{l}{\textbf{No (0)}: 360522 (77\%) , \textbf{Yes (1)}: 110403 (23\%)} \\\hline
\rowcolor{LightYellow}
\cellcolor{white}{Social connection between PR creator and the repository (H2)}  & user\_accepted\_repo & If the PR creator had a contribution accepted in the repository earlier & \multicolumn{4}{l}{\textbf{No (0)}: 198087 (42\%), \textbf{Yes (1)}: 272838 (58\%)} \\\hline
 \multirow{2}{4cm}{Track record of the PR creator (H3)} & creator\_submitted* & Number of PRs submitted by the PR creator across NPM projects & 0 & 12 & 282 & 1043 \\
 & creator\_accepted & Fraction of PRs submitted by the PR creator accepted across NPM projects & 0 & 0.64 & 0.53 & 1.00 \\\hline
\multirow{3}{4cm}{Characteristics of the PR review phase(H4)} & comments & Number of discussion comments against the PR & 0 & 2 & 3 & 11 \\
 & review\_comments & Number of code review comments against the PR & 0 & 0 & 1 & 6 \\
 & age* & Seconds between PR creation and PR closure & 231 & 100*1e3 & 651*1e3 & 2.5*1e6 \\\hline
\multirow{2}{4cm}{PR creator experience (H5)} & creator\_total\_commits* & Total number of commits made by the PR creator across git Projects & 4 & 786 & 9847 & 12,386 \\
 & creator\_total\_projects* & Total number of projects the PR creator contributed to across git Projects & 3 & 1632 & 6481 & 31,880 \\\hline
\multirow{2}{4cm}{Repository characteristics (H6)} & repo\_submitted* & Number of PRs submitted against the repository & 9 & 787 & 4787 & 30,270 \\
 & repo\_accepted* & Fraction of the submitted  PRs accepted by the repository & 0.1 & 0.70 & 0.63 & 0.91 \\\hline
 \rowcolor{LightYellow}
\cellcolor{white}{Dependency between PR creator's projects and the package (H7)} & dependency* & If any of the repositories the PR creator contributed to depend on the package & \multicolumn{4}{l}{\textbf{No (0)}: 82215 (17\%), \textbf{Yes (1)}: 388710 (83\%)} \\ \bottomrule
\end{tabular}%
}
\end{table*}

\subsection{Selection of Measures for Verifying the Hypotheses H1-H7}
We constructed the measures that could describe the factors we suspect might affect the probability of a PR being accepted from our collected dataset by consulting prior work, monitoring the discussion on different online communities, and from our experience. A detailed description of the variables is available in Table~\ref{t:vars}.

\subsubsection{Outcome Measure}
The outcome measure, i.e. the variable that we are trying to predict is whether a PR is accepted or not, and is essentially a dichotomous variable.

\subsubsection{Measures related to the technical characteristics of a PR} 
In addition to the number of commits, lines added, lines deleted, changed files in the PR, all of which are continuous variables related to its size, we also considered two dichotomous variables, describing if the PR contained any test code and whether the PR description explicitly mentioned that it contains an issue fix, as the measures that describe the technical characteristics of a PR.

\subsubsection{Measures related to the social proximity  between the PR creator and the repository to which the PR is being created} 
Following the recommendation by~\cite{tsay2014influence}, we considered a dichotomous variable, describing if the PR creator had previously created a PR against the repository which was accepted to be indicative of the social proximity  between the PR creator and the repository.

It is worth mentioning that the PR creator's association with the
repository where the PR was submitted was found to be one of the most important
predictors by~\cite{tsay2014influence}. However, we decided not to use
this variable in our final dataset, since the value of this field can
be updated retroactively (e.g. a PR creator, who had no association
with a project when first submitting a PR, might become
a \textit{member} later, and the corresponding field in the first PR
might be updated after its acceptance/ rejection), and we have no way to know the creator's
association at the time when the PR was submitted, or to verify that the affiliation wasn't updated retroactively, which would be required
to faithfully reconstruct the data as it were at the time the PR was
created. In fact, we suspect the affiliation is indeed updated retroactively, since 
we found that all PR creators with an accepted contribution to a repository have some 
association with it. We,
therefore, suspect that its importance in prior work could be due to the so
called data leakage~\cite{tu2018careful}, when 
the information leaked from the ``future'' makes prediction models misleadingly optimistic.

\vspace{-5pt}
\subsubsection{Measures related to the track record of the PR creator}
After considering the measures used by previous studies to describe the track record of the PR creator, we decided to focus on two variables that we believe adequately captures their track record: the number of PRs created by the PR creator (before creating the one under consideration) in the same ecosystem and what fraction of those have been accepted. The first measure wasn't actually used in any previous study, but we believed that it would also be an important measure and decided to include it.

\subsubsection{Measures describing the PR review phase}
We considered the following variables to be descriptive of the PR review phase: the number of comments and review comments, along with the age of the PR, which wasn't used in any previous study, but should also reflect the complexity of the evaluation process for the PR.

\subsubsection{Measures describing the experience of the PR creator}
As described in Section~\ref{s:rq}, we are looking for the specific measures describing the overall experience of the PR creator, viz. the total number of commits they created and the total number of projects they have contributed to across all projects that use git. We believe that the overall experience of the developers should be reflected in the PRs they create, for example, a developer with a considerable amount of experience in contributing to different OSS projects would likely create a good quality PR even if it is the first time they are trying to contribute to it.

\subsubsection{Measures reflecting the PRs received and accepted by a repository}
As described in Section~\ref{s:rq}, we believe the policy of a repository about accepting PRs should have an impact of PR acceptance probability, which should be captured by the following two variables: the number of PRs submitted to the repository before the submission of the one under consideration, which should reflect the popularity of the repository and how much workload the integrators might have, and the fraction of PRs that are accepted, which should be reflective of the leniency of repository towards accepting contributions.

\subsubsection{Measure describing if the PR creator depends on the package to which they are submitting a PR}
This measure is a dichotomous variable showing if any of the projects the PR creator has contributed to prior to submitting the PR under consideration depend on the repository/ package to which the PR is submitted.

\begin{figure}[!t]
\centering
\vspace{-10pt}
\includegraphics[width=\linewidth]{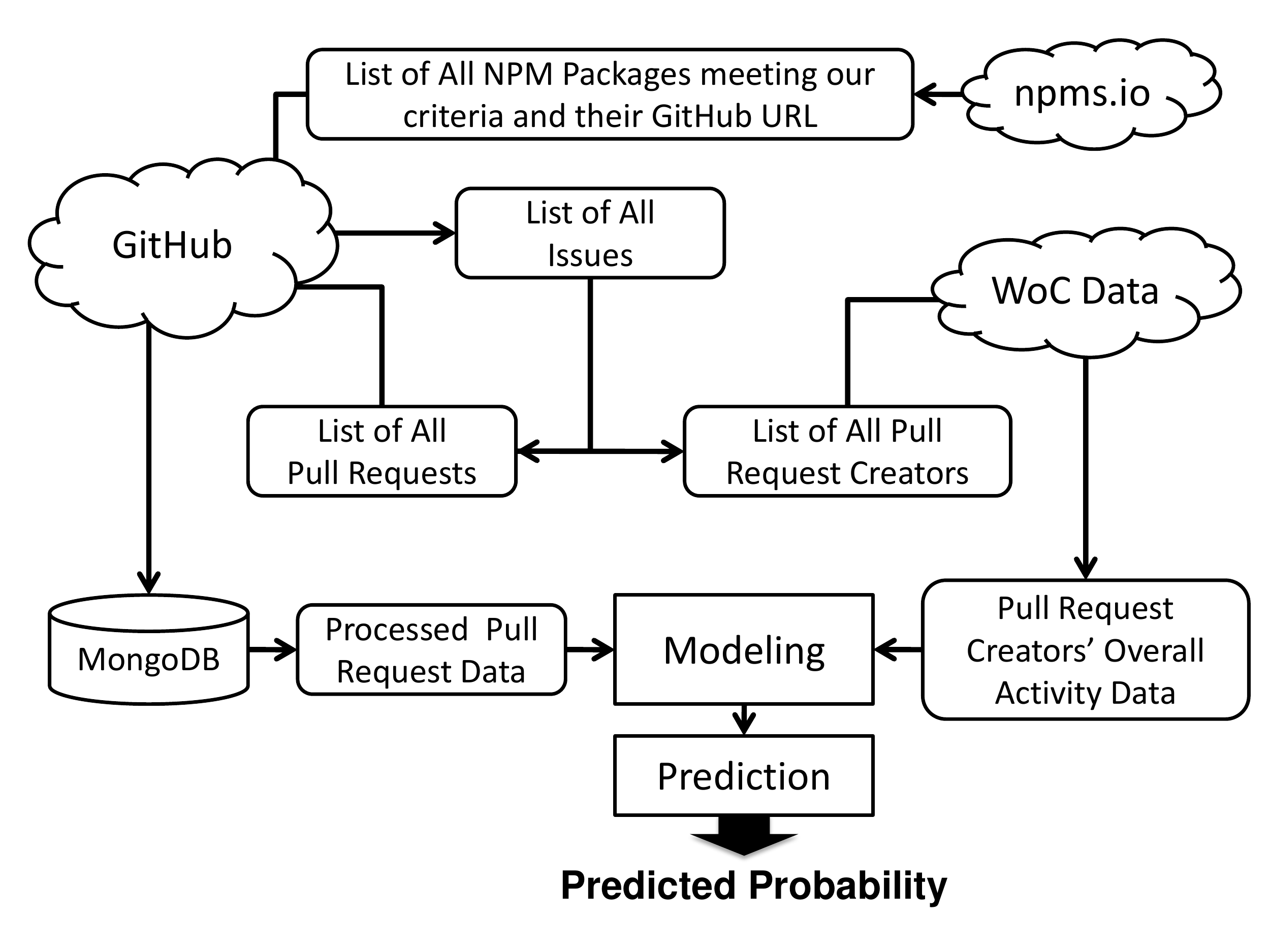}%
\vspace{-10pt}
\caption{The Data Collection and Modeling Architecture.}
\label{fig:flow}
\vspace{-10pt}
\end{figure}

\subsection{Data Collection}
To be able to identify the PRs that adhere to our criteria, as mentioned above, we first needed the list of all NPM packages that have more than 10,000 downloads per month and a GitHub repository with at least 5 PRs. This was obtained from the \texttt{npms.io} website using their API.~\footnote{https://api.npms.io/v2/package/[package-name]} The associated GitHub repository URLs were collected from the metadata information of these packages, which was obtained by using a ``follower" script, as described in NPM's GitHub repository.~\footnote{https://github.com/npm/registry/blob/master/docs/follower.md} After filtering for our criteria that the NPM package must have more than 10,000 monthly downloads (since January, 2018) and an active GitHub repository, we were left with 4218 different NPM packages. Next, we needed the list of all PRs for these NPM packages. To obtain this, we first collected all the issues associated with these NPM packages, since GitHub considers PRs as issues, and then identified the issues that have an associated patch, i.e. the ones that are PRs. The list of all issues for the packages was obtained using the GitHub API for issues\footnote{https://developer.github.com/v3/issues/}, using the \texttt{state=all} flag. We identified 483,988 PRs for all the issues for these 4218 packages. It is worth mentioning here that sometimes more than one NPM package can have the same associated GitHub repository, e.g. all TypeScript NPM packages (starting with ``@types/'', like @types/jasmine, @types/q, @types/selenium-webdriver etc.) refer to GitHub repository ``DefinitelyTyped/DefinitelyTyped''. To avoid double-counting and further confusion, we saved the issues keying on the repository instead of the package name, though we also saved the list of packages associated with a repository. We found that there are 3601 unique repositories associated with these 4218 packages. We further filtered this dataset to only have the repositories that had at least 5 PRs, and found that 2740 repositories, associated with 3349 NPM packages, match this criteria. Then, we obtained the data on all the PRs for these repositories from GitHub using the API.\footnote{https://developer.github.com/v3/pulls/} This data was stored in a local MongoDB database. We used a Python script to extract the data from this database and process it into a CSV file that was used for modeling. We further filtered out all PRs that were not marked ``closed'', since we are only interested in looking at the already resolved PRs. 

The data on the PR creators' overall activity across all projects that use git were obtained from the World of Code (WoC) data~\cite{woc19}. WoC is a prototype of an updatable and expandable infrastructure to support research and tools that rely on version control data from the entirety of open source projects that use git. The data in WoC is stored in the form of different types of maps between different git objects, e.g. there are maps between commit authors and commits, commits and projects etc. The detailed description of this dataset is available in~\cite{woc19} and the project website.~\footnote{\url{https://bitbucket.org/swsc/overview/src/master/}}
Specifically, we used this dataset to compile the profiles of PR authors. Since the author ID used in WoC is different from the GitHub ID of developers (WoC author IDs comprise the name and email address of the developers), we
identified the PR authors by first obtaining the commits they included in their
PRs, and then by identifying the author IDs for these commits in WoC
using commit to author maps. Then we identified all the remaining
commits for these authors using the author to commit map. That full set
of commits for each author was used to count the number of projects they contributed to. To construct the relevant measures for the PR acceptance
prediction, we only used the commits made by the PR author prior to
the creation of the PR, thus reconstructing the state of affairs as
it existed at the time of PR creation.

\subsection{Variable Construction} 
As shown in Table~\ref{t:vars}, we listed 17 different measures related to the hypotheses we proposed in Section~\ref{s:rq}. The data for 6 of those measures, \textit{additions, deletions, commits, changed\_files, comments,} and \textit{review\_comments}, were directly obtained from the data collected from GitHub API. The data on two measures, \textit{creator\_total\_commits} and \textit{creator\_total\_projects}, were obtained directly from the WoC dataset. To calculate the \textit{dependency} measure, we considered all the commits made by the PR creator and, using those, find out which projects they have contributed to. Then we check the dependencies of all those projects to see if any of them list the package as a dependency.

Calculation for the \textit{age} variable was relatively straightforward, we simply  counted the seconds between the time of PR creation and the time of PR closure. For determining whether the PR contained any test code or mentions fixing an issue, we looked at its description and populated the measures \textit{contain\_test\_code} and \textit{contain\_issue\_fix} based on whether the description of the submitted PR mentioned including test code (we looked for the phrase ``test code'' and a few of its variations in the description) and fixing an issue (we checked if the description has one of the words signifying ``fix'', e.g. ``fix'', ``resolve'' etc. and an issue, e.g. the word ``issue'', the symbol ``\#'' followed by a number etc.).

For calculating the remaining variables, we sorted our whole dataset by PR creation times and counted the cumulative number of PRs created by each PR creator and against each repository and kept track of the fact of whether they were accepted or not. Using this cumulative data, we calculated the values for the variables \textit{creator\_submitted, creator\_accepted, repo\_submitted, repo\_accepted,} and \textit{user\_accepted\_repo}.

Since the values of all continuous variables except \textit{creator\_accepted} and \textit{repo\_accepted} were found to be heavily skewed, we converted them into log scale for modeling purposes.

\vspace{-10pt}
\subsection{Analysis Method}
We used logistic regression for verifying which variables have a significant impact on PR acceptance probability, and used Random Forest method for measuring the predictive performance of our model and ranking the measures by importance. The variations of PR acceptance probability with the values of these measures were identified using partial dependence plots~\cite{milborrow2019plotting}.

\vspace{-10pt}
\section{Results}\label{s:result}

\subsection{General Background}\label{ss:gen}

Our study focused on 3349 NPM packages (2740 unique GitHub repositories) with more than 10,000 monthly downloads since January, 2018, an active GitHub repository, and at least 5 PRs created against them. We collected 470,925 pull-requests, which were created by 79,128 unique GitHub users, and were evaluated by 3633 unique integrators.  

Of these PRs, 290,058 (61.6\%) were accepted (merged into the codebase), which were created by 47,099 unique GitHub users (59.5\% of all PR creators). 87 repositories (3\% of the ones under consideration), which had a total of 981 PRs created against them, didn't accept any of the PRs. Conversely, 124 (4.5\% of total) repositories, who received 4230 PRs in total, accepted all of them. 

\vspace{-10pt}
\begin{table}[ht]
\caption{Regression Model showing the significance of the measures listed in Table~\ref{t:vars} in explaining PR acceptance. P-values less than 0.0001 are shown as 0. Measures not found to be significant are highlighted in Red. Dichotomous variables are shown in blue.}
\label{t:regression}
\resizebox{\linewidth}{!}{%
\begin{tabular}{p{3cm}lcr}
\hline
\textbf{Underlying Factors (Hypothesis)} & \multicolumn{1}{c}{\textbf{Measure }} & \textbf{Coefficient $\pm$ Std. Error}  & \textbf{p-Value} \\ \hline
& (Intercept) & 0.3830 $\pm$ 0.0255 & 0 \\\hline
PR Characteristics (H1) & additions & -0.0168 $\pm$ 0.0030 & 0 \\
\rowcolor{LightRed}
\cellcolor{white} & deletions & -0.0010 $\pm$ 0.0029 & 0.7332 \\
& commits & -0.2475 $\pm$ 0.0076 & 0 \\
 & changed\_files & 0.0219 $\pm$ 0.0073 & 0.0026 \\
& \textcolor{blue}{contain\_issue\_fix:1} & 0.0338 $\pm$ 0.0077 & 0 \\
\rowcolor{LightRed}
\cellcolor{white} & \textcolor{blue}{contain\_test\_code:1} & 0.1046 $\pm$ 0.1236 & 0.3976 \\\hline
Social connection between PR creator and the repository (H2)
& \textcolor{blue}{user\_accepted\_repo:1} & 0.7921 $\pm$ 0.0111 & 0 \\\hline
\multirow{2}{3cm}{Track record of the PR creator (H3)} & creator\_submitted & -0.1371 $\pm$ 0.0028 & 0 \\
& creator\_accepted & 1.3590 $\pm$ 0.0128 & 0 \\\hline 
\multirow{3}{3cm}{Characteristics of the PR review phase (H4)} & comments & -0.4519 $\pm$ 0.0054 & 0 \\
& review\_comments & 0.2785 $\pm$ 0.0059 & 0 \\
& age & -0.2100 $\pm$ 0.0015 & 0 \\\hline 
\multirow{2}{3cm}{PR creator experience (H5)} & creator\_total\_commits & 0.0115 $\pm$ 0.0027 & 0  \\
& creator\_total\_projects & 0.0256 $\pm$ 0.0023 & 0 \\\hline 
\multirow{2}{3cm}{Repository characteristics (H6)}
& repo\_submitted & 0.0071 $\pm$ 0.0017 & 0 \\
& repo\_accepted & 3.3468 $\pm$ 0.0174 & 0 \\\hline 
Dependency between PR creator's projects and the package (H7) & \textcolor{blue}{dependency:1} & 0.0317 $\pm$ 0.0099 & 0.0014 \\ \hline
\end{tabular}%
}
\vspace{-10pt}
\end{table}

\subsection{Testing the significance of the measures}
In order to find out which variables have a significant effect on PR acceptance, we used a logistic regression model, the result of which is presented in Table~\ref{t:regression}. We checked the variance inflation factors (VIF) for this model to ensure there is no multicollinearity problem and found the maximum value of VIF to be 3.1, which is within acceptable threshold. Two out of the total 17 measures used in the model as predictors were found to be insignificant, both of which are related to the technical characteristics of a PR.

\subsubsection{Examining H1}
The measure showing whether the PR included test code was found to be significant in~\cite{tsay2014influence}, but turned out to be insignificant for our dataset. The size of a PR was also reported to be significant in~\cite{tsay2014influence,weissgerber2008small}, who looked at the number of lines changed. We described the lines changed by two different variables: number of lines added and number of lines deleted. While the number of lines added was found to be a significant predictor, the number of lines deleted was found to be insignificant for our dataset. Two other variables we hypothesized to be descriptive of the PR's technical characteristics: the number of commits in the PR and whether the PR description explicitly mentioned fixing an issue were both found to be significant predictors. Explicit mention of an issue fix in the PR description was found to increase the chance of a PR being accepted. The number of lines added and the number of commits negatively affect the PR acceptance probability, which suggests that, in general, smaller patches have a higher chance of being accepted, and supports the findings of ~\cite{tsay2014influence,weissgerber2008small}. However, we can see from Table~\ref{t:regression} that the number of changed files seems to have a positive effect on PR acceptance probability, which is contrary to what was reported in~\cite{tsay2014influence}. Upon further inspection, the actual scenario turned out to be a bit more complex, which is discussed in detail in Section~\ref{ss:partial}.
Although a couple of measures related to this factor were insignificant, the technical characteristics of a PR as a whole were seen to have a significant impact, though only some of the relationships reported previously could be replicated in our case.  

\subsubsection{Examining H2}
We found the measure describing the social proximity  between the PR creator and the repository to be significant. Having a contribution accepted in the project earlier had a positive influence on PR acceptance probability, similar to what was reported by~\cite{tsay2014influence}. This finding increase the generalizability of the social proximity as an important predictor of PR acceptance.

\subsubsection{Examining H3}
The number of PRs submitted by the PR creator (before the one under consideration) and the fraction of those PRs that were accepted both proved to be significant predictors for explaining PR acceptance probability. While the total number of PRs submitted earlier had a negative influence, possibly due to the presence of relatively inexperienced PR creators in the dataset whose submissions aren't accepted, a higher fraction of accepted PRs had a strong positive influence on the probability of PR acceptance, showing the importance of a good track record, as was also reported in~\cite{prioritizer}.

\subsubsection{Examining H4}
The variables describing the characteristics of the PR review phase: the number of discussion comments, the number of code review comments, and the age of the PR were all found to be significant predictors. The number of discussion comments was observed to have a negative influence on PR acceptance probability, similar to what was observed by~\cite{tsay2014influence}. Although the actual situation might be a bit complex, as described in Section~\ref{ss:partial}, the number of code review comments was found to have a positive influence on PR acceptance overall. Therefore, the general statement made by~\cite{tsay2014influence} that highly discussed contributions are less likely to be accepted seems to come with a caveat: it is true if we are referring to the discussion comments, but false if we talk about the code review comments. The age of a PR seems to have a negative influence on the acceptance probability in general, indicating that good quality PRs are accepted pretty quickly in the NPM ecosystem. 

\subsubsection{Examining H5}
The overall experience of a developer was seen to have a significant effect on the probability of their PR contributions being accepted, with the probability of acceptance increasing with the number of commits made by the PR creators and the number of projects they have contributed to.
While this appear intuitive, we hope that other studies on different ecosystems would confirm these findings.

\subsubsection{Examining H6}
The characteristics of the repository were also found to have a significant impact on PR acceptance probability. The number of PRs submitted to a repository as well as the fraction of PRs accepted by it had a positive impact on PR acceptance probability. The faction of PRs accepted was the most influential variable we had (it had the highest z-score). This observation leads us to believe that our hypothesis, that the leniency of a repository towards accepting PRs has a significant impact on PR acceptance, holds. It further strengthens the credibility of the assumption that contextual factors, especially ones that are repository-specific, might play a substantial role.

\subsubsection{Examining H7}
The hypothesis that if any of the projects the PR creator contributed to has a dependency on the package to which the creator submitted a PR, it has a higher chance of being accepted also holds according to our model, as can be observed from Table~\ref{t:regression}.

Overall, we found that all of the null hypotheses, denoted by H1.0-H7.0, (null hypothesis is that the postulated factors have no effect on PR acceptance) corresponding to the ones we presented (H1-H7) were rejected, which allows us answer the first Research Question we posed. Although we can see the direction of the relationships between the variables and PR acceptance probability may not always be intuitive as shown in Table~\ref{t:regression}, that aspect is examined in more detail later (Section~\ref{ss:partial}). While, generally, we got results consistent with prior literature, there were a number of exceptions suggesting that further studies may be 
needed to resolve these inconsistencies.

\hypobox{Answering RQ1: All of the null hypotheses (H1.0-H7.0) we posed were rejected on our dataset, indicating the factors found to be significant in earlier studies (H1-H4) as well as the ones we proposed (H5-H7) have significant influence on PR acceptance.}

\subsection{Relative Importance of the measures in predicting PR acceptance}

\begin{figure}[!t]
\centering
\vspace{-10pt}
\includegraphics[width=0.9\linewidth]{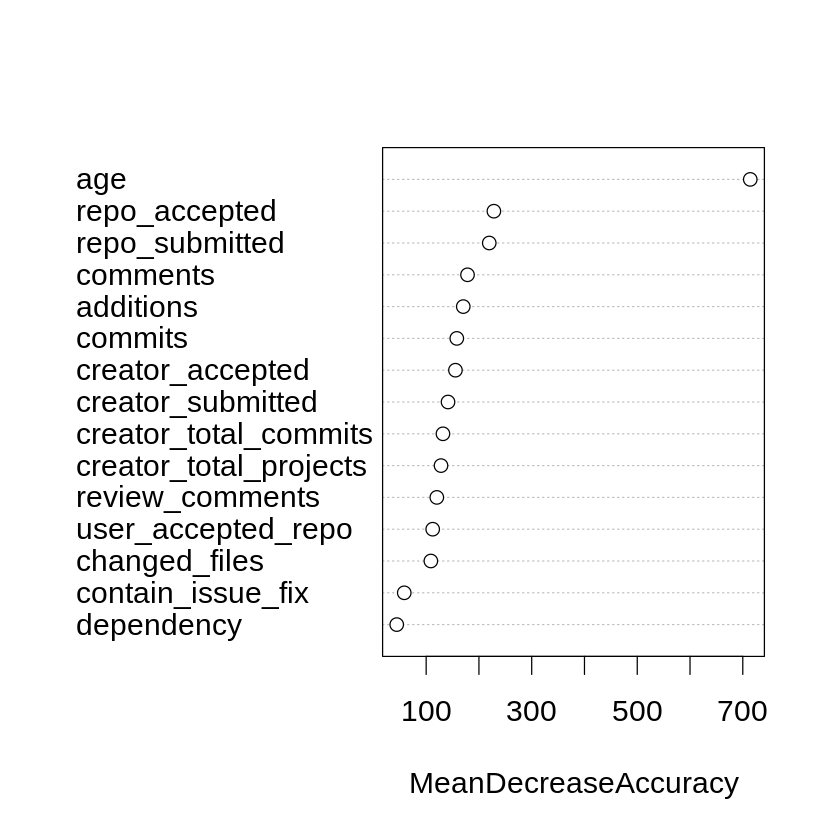}%
\vspace{-10pt}
\caption{Variable Importance Plot from the Random Forest model for predicting PR Acceptance}
\label{fig:vimp}
\vspace{-10pt}
\end{figure}

To gauge their predictive power and determine their relative importance for predicting if a PR will be accepted
we created a Random  Forest model with the 15 predictors found to be significant from our earlier analysis. Although our observations are independent of each other, there is an underlying element of time in the whole dataset. Therefore, to ensure that there is no data leakage concerns, we decided to divide the data such that the model is trained on 70\% of the PRs that are created before the rest, and we tried to predict the acceptance of the remaining 30\% of the PRs. We repeated the process 1000 times to ensure there is no significant variation in performance and the relative importance of the predictors.

Our model achieved an AUC of 0.94, with the values of sensitivity and specificity being 0.69 and 0.76 respectively. The variable importance plot, which shows the most common order of the relative importance of the variables in terms of mean decrease in accuracy,  is presented in Figure~\ref{fig:vimp}. This ordering lets us understand the relative importance of different measures in predicting PR acceptance and answer RQ2. We also observed that the top three measures in terms of importance were among the 7 measures we introduced in this study (see Table~\ref{t:vars}).

\hypobox{Answering RQ2: The age of the PR is the most important variable for predicting PR acceptance, followed by the two repository characteristics measures and the measures related to the size of the PR. Measures describing the review phase (other than age) and the track record and experience of the PR creator also turn out to be relatively important. Comparatively, measures related to the social proximity  between the creator and the repository, the measure of PR characteristics not related to its size, and the one describing if there is a dependency between the creator's projects and the package turn out to have lower importance in predicting PR acceptance. }

\begin{figure*}[!htb]
\caption{Partial Dependence plots for the 15 measures from the Random Forest model for predicting PR acceptance probability. The plots were generated using ``randomForest''~\cite{randomForest} package in R, and smoothed for ease of interpretation with ``ggplot2''~\cite{ggplot2} package in R using generalized additive models (GAM).}
\label{t:pplot}
\resizebox{0.9\linewidth}{!}{%
\begin{tabular}{llll}
\includegraphics[width = \linewidth]{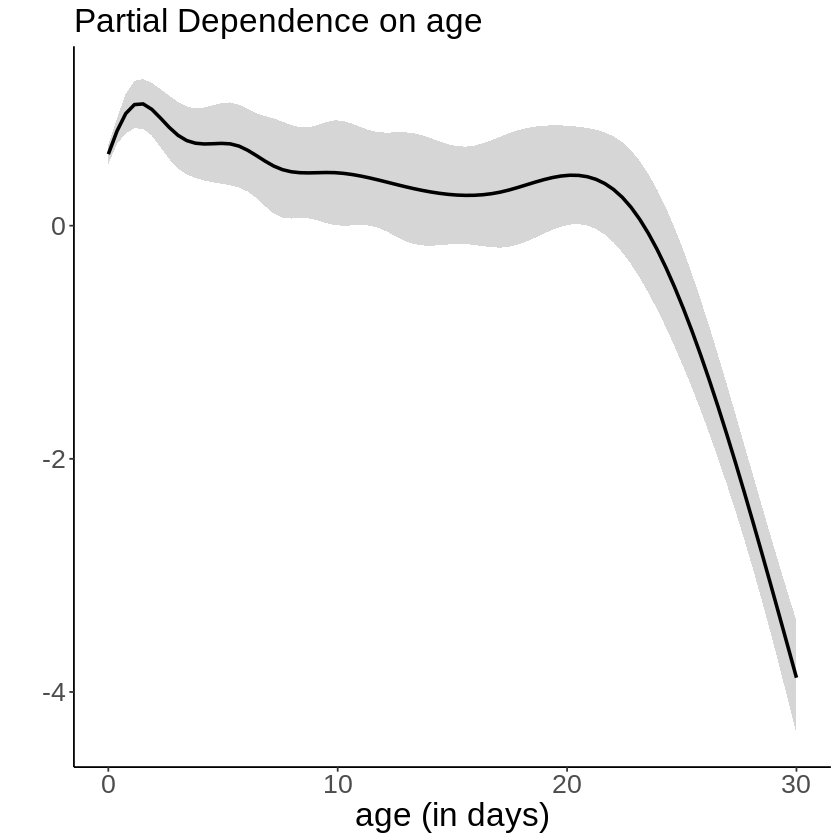} &
\includegraphics[width = \linewidth]{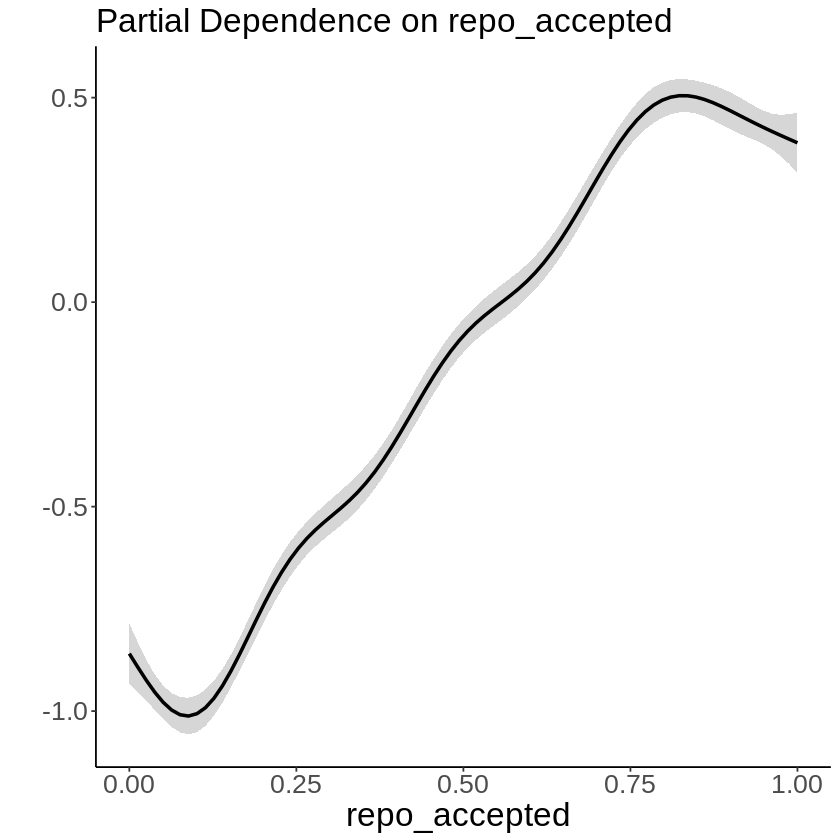} &
\includegraphics[width = \linewidth]{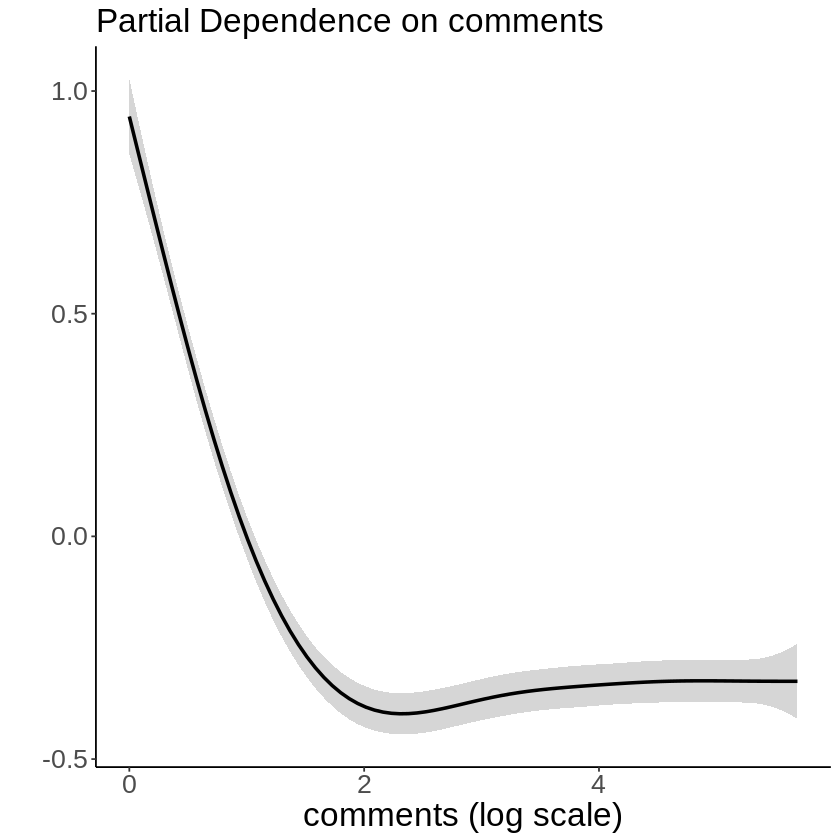}&
\includegraphics[width = \linewidth]{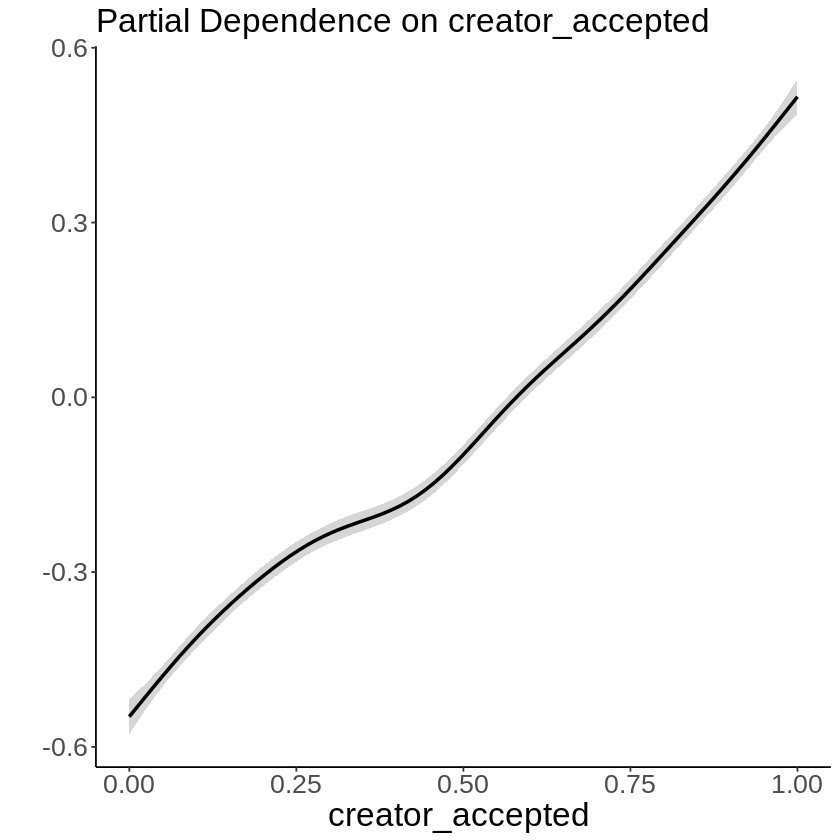} \\
\includegraphics[width = \linewidth]{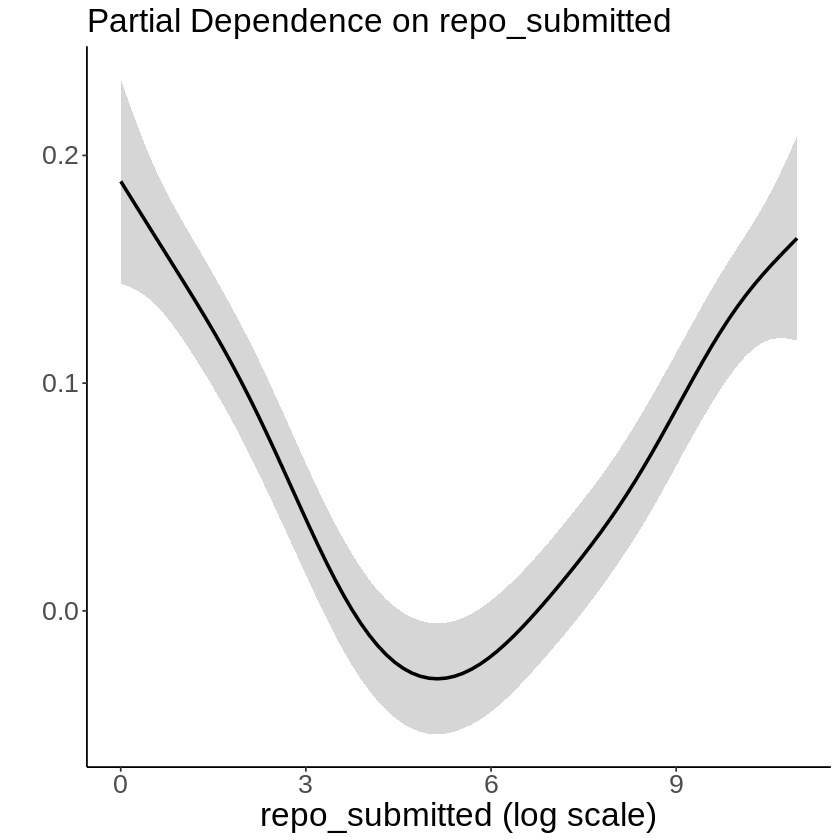} &
\includegraphics[width = \linewidth]{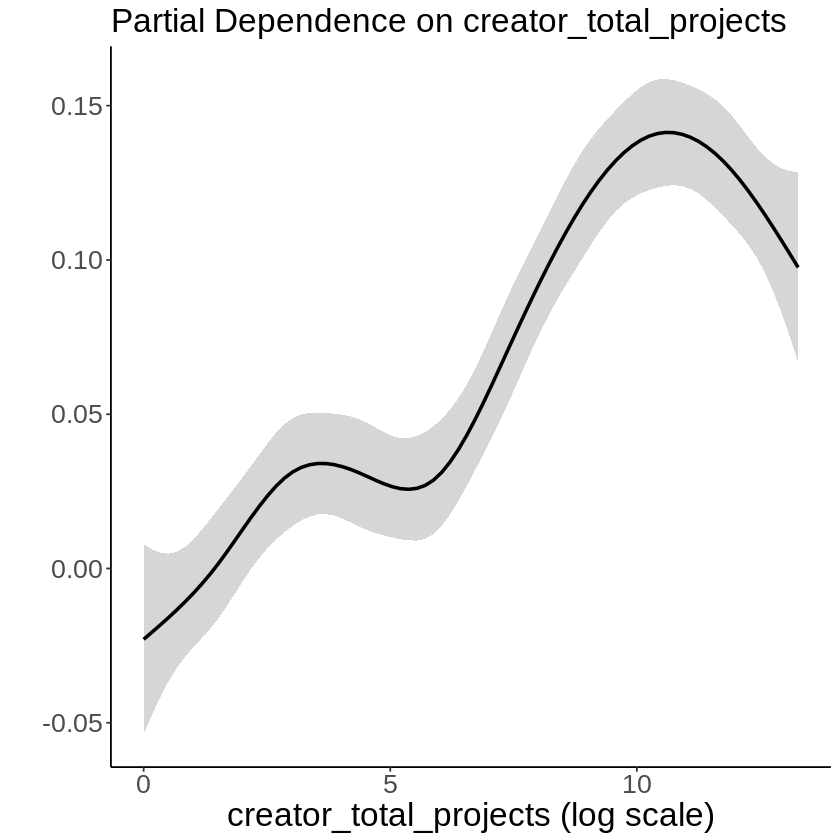} &
\includegraphics[width = \linewidth]{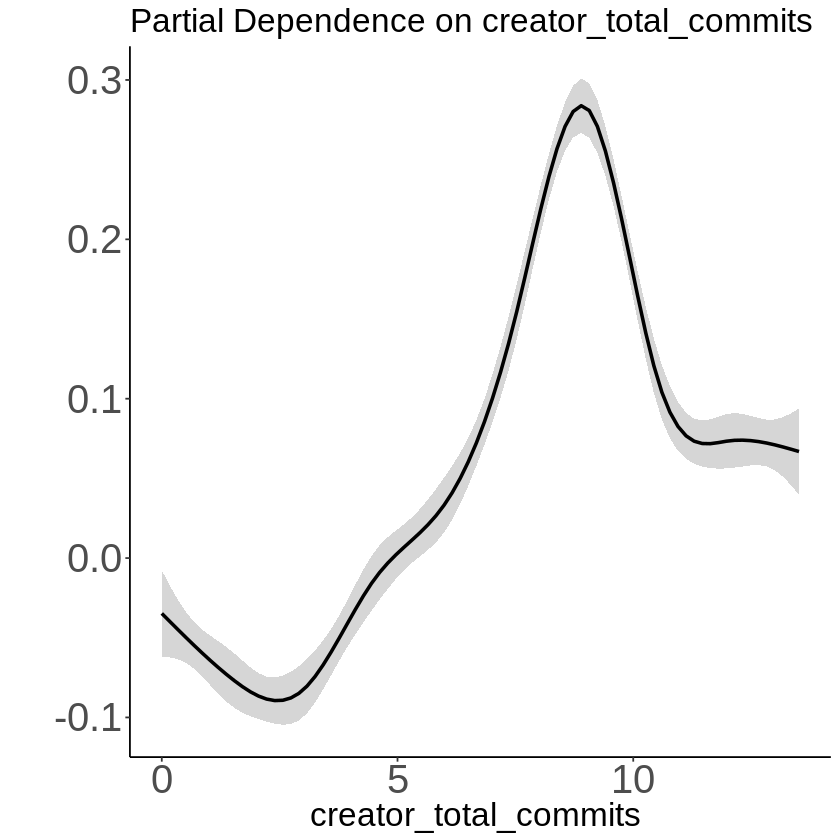} &
\includegraphics[width = \linewidth]{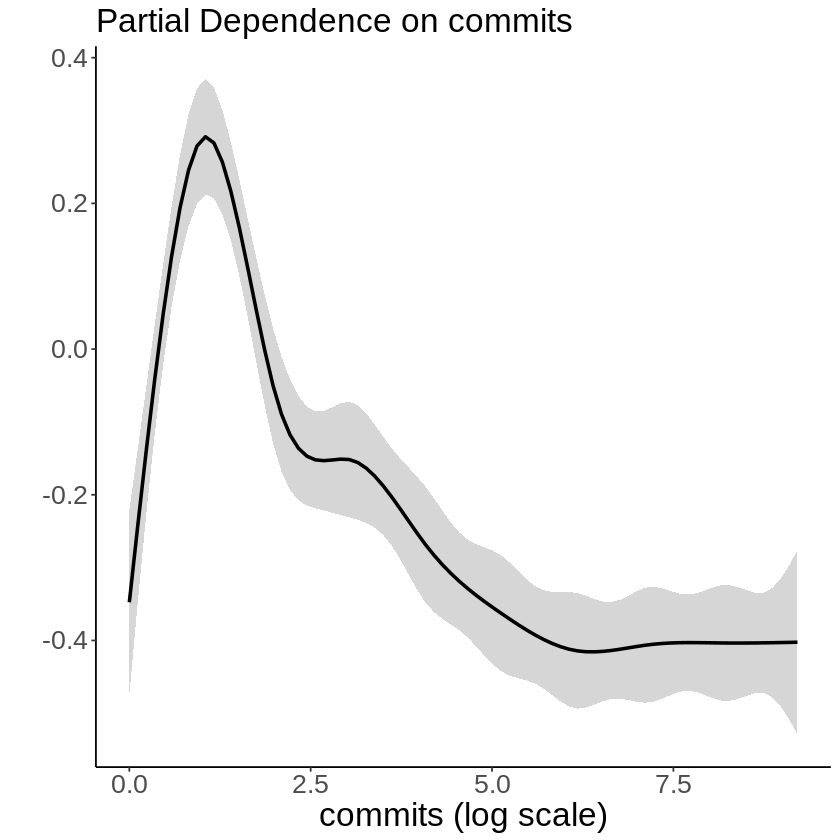} \\
\includegraphics[width = \linewidth]{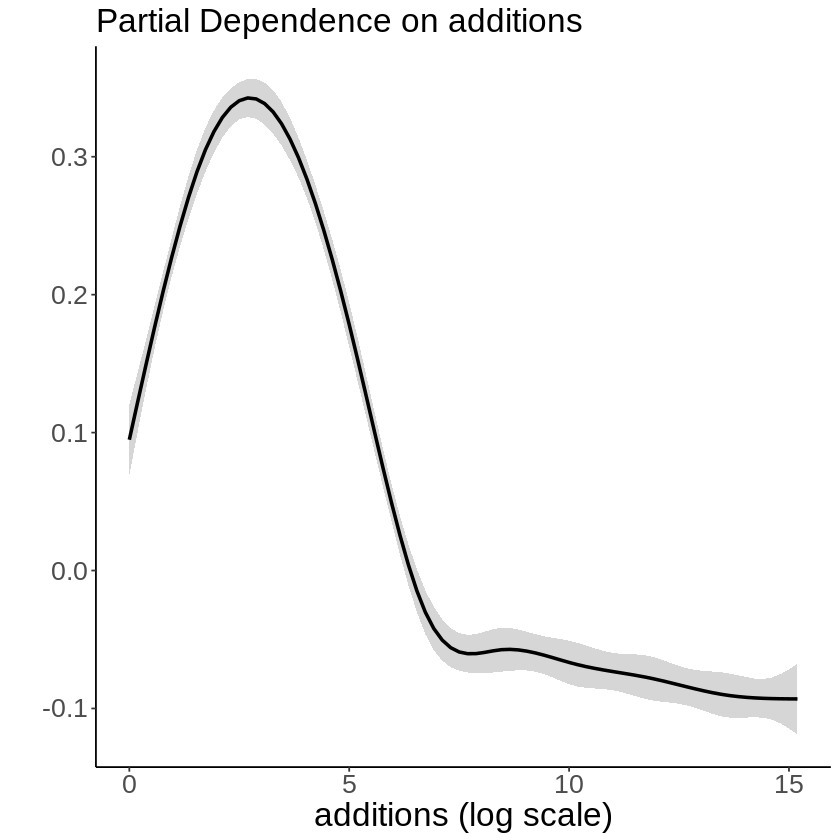} &
\includegraphics[width = \linewidth]{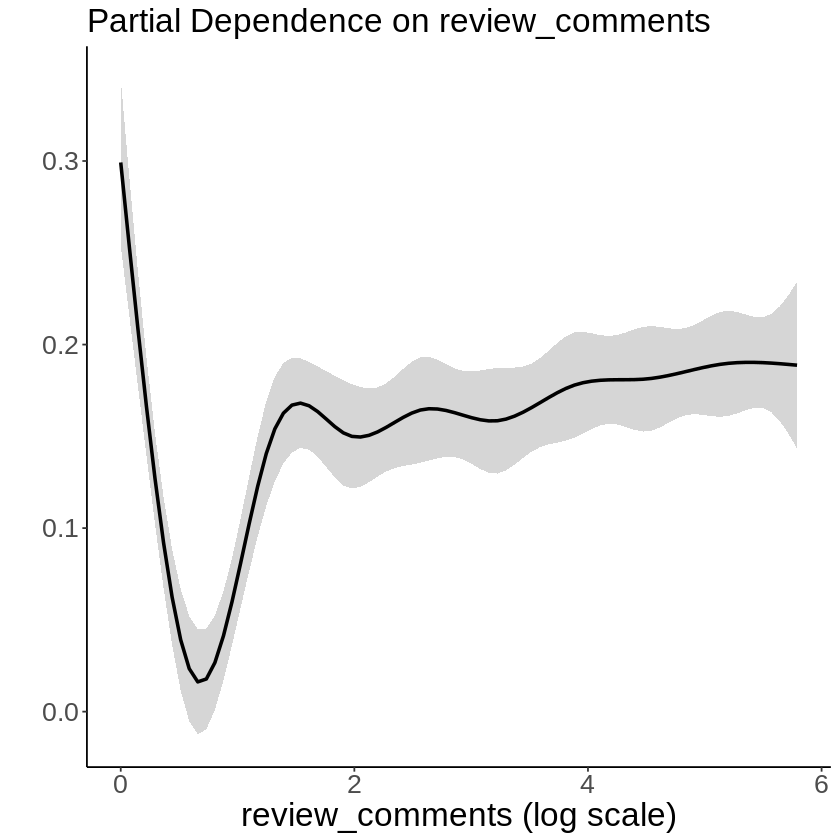} &
\includegraphics[width = \linewidth]{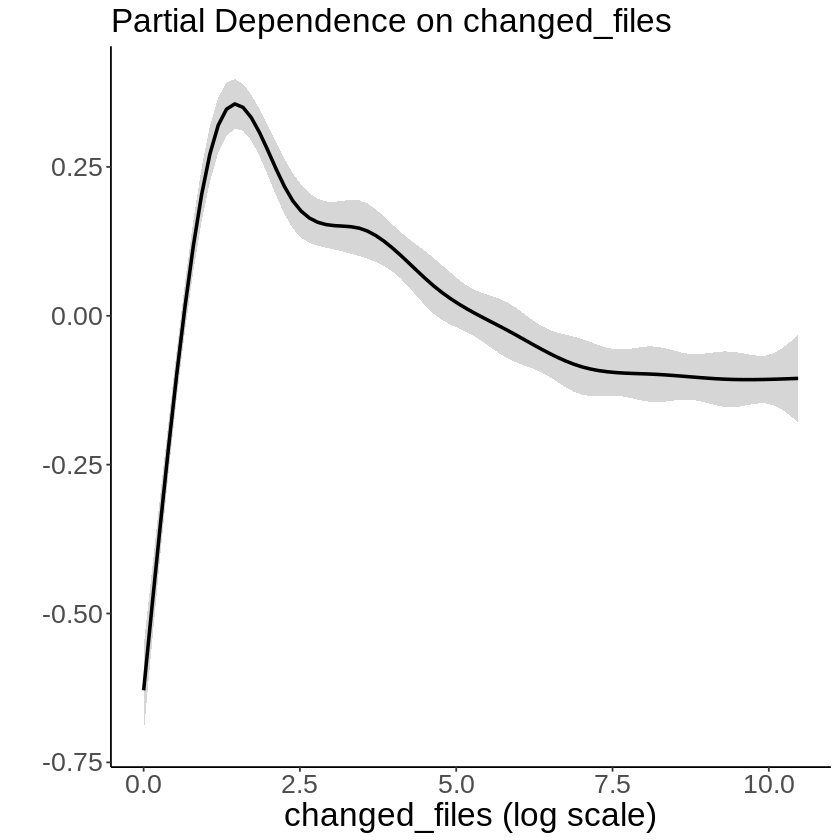} &
\includegraphics[width = \linewidth]{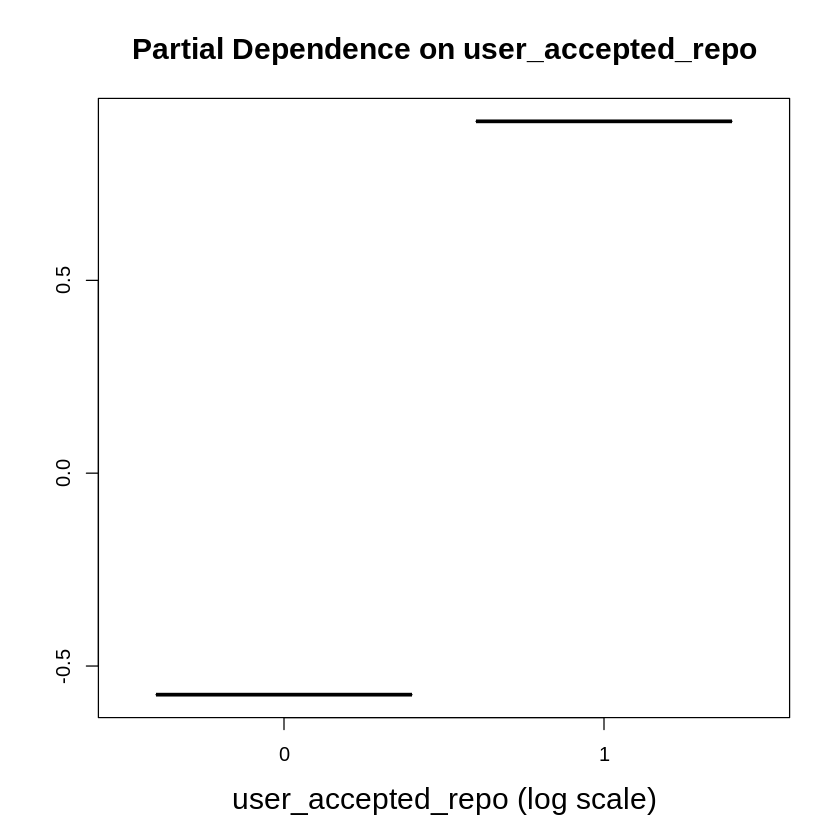} \\
\includegraphics[width = \linewidth]{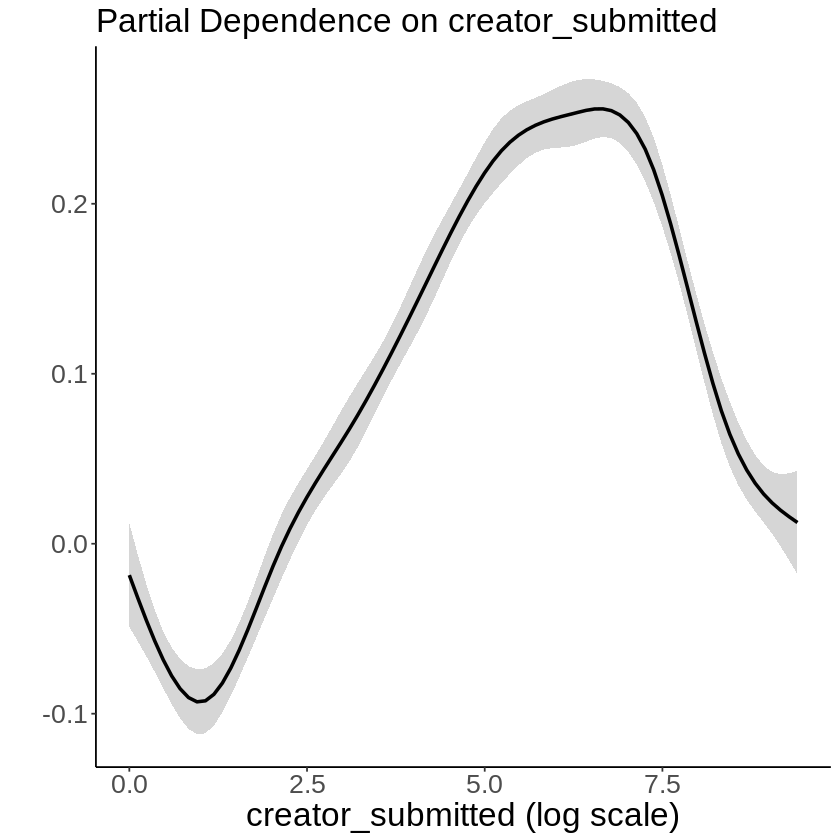} &
\includegraphics[width = \linewidth]{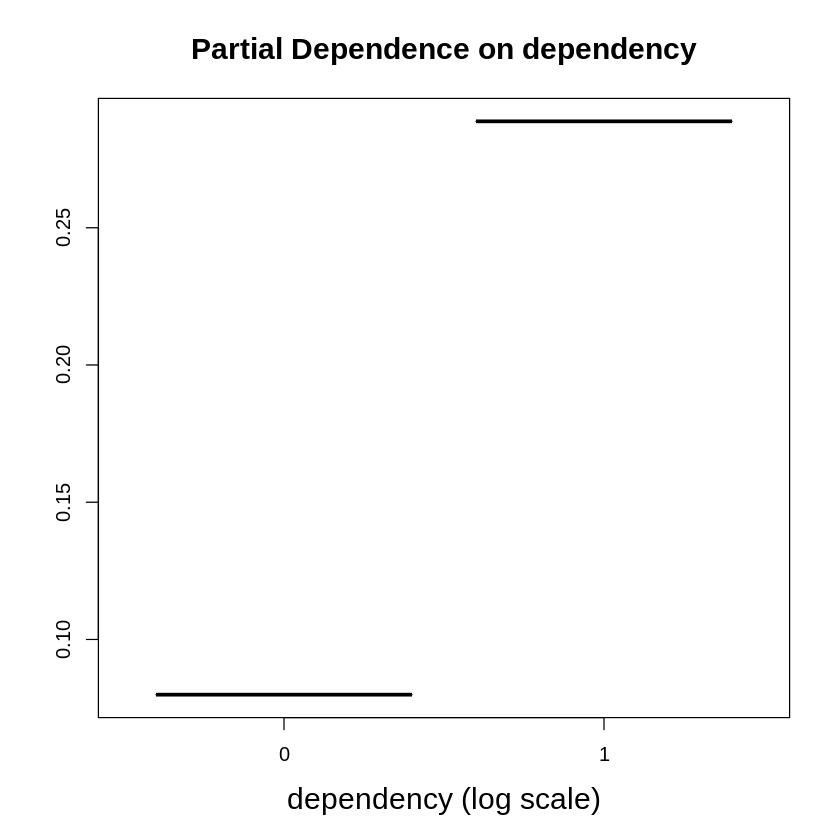} &
\includegraphics[width = \linewidth]{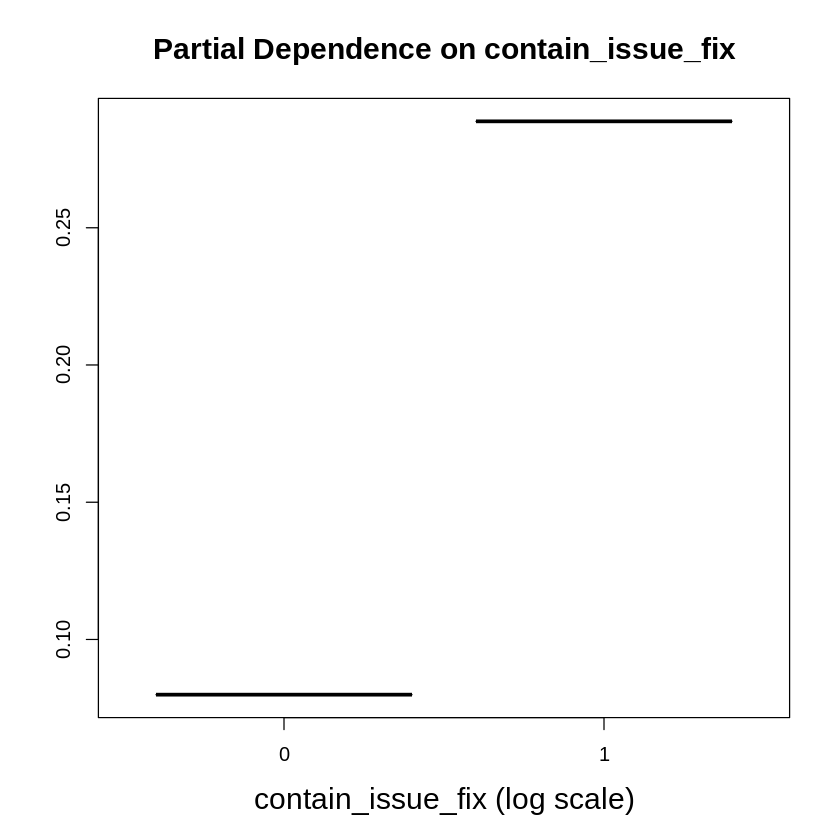} &

\end{tabular}%
}
\vspace{-10pt}
\end{figure*}

\subsection{Variation of the PR Acceptance Probability with the Predictors' values}\label{ss:partial}
To identify how the probability of PR acceptance varies with the values of the 15 measures found to be significant, we generated partial dependence plots,~\cite{milborrow2019plotting,friedman2001greedy} which are graphical depictions of the marginal effects of these variables on the probability of PR acceptance, from the Random Forest model we created. 
In the X axes of a plots we have the values of the independent variables and the Y axes of the plots show the relative logit contribution of the variable on the class probability~\cite{friedman2001greedy} (probability that a PR was merged, in our case) from the perspective of the model, i.e. negative values (in the Y-axis) mean that the positive class is less likely (i.e. it is less likely that a PR would be accepted, in our case)  for that value of the independent variable (X-axis) according to the model and vice versa. These plots can shed light into the dynamics of PR creation and acceptance, and would be helpful for both the PR creators and the integrators for understanding how to improve the quality of PRs being submitted and accepted. 

The resultant plots for the different measures are presented in a tabular format in Figure~\ref{t:pplot}. We generated the plots with the ``randomForest'' package in R. However, we observed that the plots are not entirely smooth for a few of the measures, so, in order to be able to interpret the results better, we decided to smooth the plots using generalized additive models, which was achieved by using the \texttt{geom\_smooth} function of the ``ggplot2'' package in R with the option \texttt{method = ``gam''}. The plots in Figure~\ref{t:pplot} are arranged by decreasing importance, as observed from Figure~\ref{fig:vimp}.

\subsubsection{Variation with PR age}
As observed from Figure~\ref{t:pplot}, the probability of a PR being accepted drops steadily with time. We also observe a catastrophic drop in PR acceptance probability as it gets older than 20 days. This suggests that the PR integrators in the NPM ecosystem tend to be quite responsive and efficient in handling PRs and older PRs may even not be considered. It may also be reflective of the rapid development in NPM ecosystem which makes it harder for the older PRs to get merged with the main development branch. 

\subsubsection{Variation with fraction of PRs accepted by a repository}
We see that the repositories that accept a larger fraction of the PRs submitted against them are more likely to accept PRs in the future, which could indicate a more lenient policy of PR acceptance and/or the integrators' willingness to accept contributions from other developers. However, we see that repositories that have very high or very low acceptance rate seem to deviate from this trend, but we suspect that this is due to the ``cold start'' problem: as a repository just starts getting PRs, the fraction of those they accept or reject changes the value of this measure dramatically. 

\subsubsection{Variation with the number of discussion comments}
We observe that the probability of PR acceptance steadily drops as the number of discussion comments increases, and, as the number of comments go beyond 7 (2.08 in the partial dependence plot in Figure~\ref{t:pplot}), the drop in probability gets saturated. This indicates that if the value added by a PR is ``obvious'', it is more likely to be accepted.

\subsubsection{Variation with the fraction of PRs created by a PR creator that are accepted}
We observe that the probability of a PR being accepted increases steadily with the value of this measure, which highlights that a good track record of a PR creator has a strong positive influence on the probability of their PRs being accepted. 
\vspace{-5pt}
\subsubsection{Variation with total number of PRs submitted against a repository}
We observe from the partial dependence plot of this measure that repositories that either have a small or a large number of PRs created against them are more likely to accept one compared to the repositories that get a moderate number of PRs. This may be because for the repositories that get a smaller number of PRs, it is easier for them to evaluate those requests and work with the PR creators to get the contributions accepted. On the other hand, repositories that get a large number of PRs tend to be quite large themselves and have support from a good number of developers, making it easier for them to evaluate PRs submitted against them. It is also possible that many of them are very open to accepting contributions, which is known to many of the PR creators as well, which makes them more willing to submit PRs to these repositories. 

\subsubsection{Variation with the total number of projects a PR creator contributed to}
We observe that having contributed to a larger number of projects steadily increase the chance of a PR creator's contributions being accepted, which is likely because developers who have contributed to a larger number of projects tend to be more experienced, more knowledgeable about the different requirements of different projects, and, in general, tend to create better PRs. However, a trend reversal is observed for creators who contributed to over 5000 projects. The reason for the trend reversal wasn't clear to us at first, so we investigated those cases further, and found that these developers tend to be bots (e.g. greenkeeper bot), not humans, which explains why the PRs created by them have a relatively lower chance of being accepted, since these bots do not gain experience and improve in the same way as human developers. 

\subsubsection{Variation with the total number of commits by a PR creator}
The probability of a PR being accepted tends to increase with the total number of commits created by a PR creator, however, we observe a trend reversal for PR creators with an extremely large number of commits. The reason for an increase in PR acceptance probability with a larger number of commits is quite straightforward: the PR creator is more experienced, so the code contributions they make tend to be of higher quality. The reason for the trend reversal for creators with a very high number of commits (over 10,000) is, once again, because most of them actually are bots.

\subsubsection{Variation with the number of commits in the PR}
We observe from the partial dependence plot that PRs with very few commits have a low chance of getting accepted, since those might not have a significant amount of contribution. Initially, the probability of PR acceptance increases rapidly as the number of commits increase. However, after reaching a peak at around 2 commits, the probability of acceptance starts dropping quickly. The drop rate slows down for PRs with over 10 commits  and gets saturated for the ones with over 300 commits. 

\subsubsection{Variation with the number of lines added in the PR}
Similar to what we observe for the commits, the PRs with very few lines added are less likely to be accepted. The probability rises to a peak for the PRs with around 20 lines added, starts dropping steadily until we reach PRs with around 400 lines added, and the rate of decrease in the chance of a PR being accepted keeps dropping slowly after that.

\subsubsection{Variation with the number of code review comments for the PR}
The partial dependence plot shows that the probability of acceptance is high for the PRs with no code review comments, the value of the acceptance probability takes a plunge for the ones with just a single code review comment (likely the comment clarifies why it couldn't be accepted), and shows a continuous moderate increase for the ones with more than 1 code review comments. This potentially reflects the fact that some PRs may require more discussion due to their complexity or impact, but are otherwise of high quality. 

\vspace{-10pt}
\subsubsection{Variation with the number of changed files in the PR}
As with the other measures related to the size of the PR, we see a steady increase in the probability of PR acceptance for the PRs with up to 4 changed files, and it shows a constant decrease after that. It is worth mentioning that most ($\sim80\%$) of the PRs in our dataset had less than 4 changed files, which is likely why we saw a positive regression coefficient for this variable in our logistics regression model (Table~\ref{t:regression}).

\subsubsection{Variation with the number of PRs submitted by the PR creator}
The variation of the PR acceptance probability with the number of PRs submitted by a PR creator is a bit complex, and we have to keep in mind that our dataset has multiple entries corresponding to a PR creator, one for each of their submitted PRs, to fully comprehend the dependence. The initial peak in the dependence plot most likely corresponds to the ``cold start'' problem, i.e. when the PR creators with a high amount of skill submit their PRs for the first time and it gets accepted. The following trough reaches its lowest point at around 10 PRs submitted by the PR creator, and then we see the experience gained by the PR creators adding up, increasing the probability of their PRs being accepted. The probability reaches its peak at around 100 PRs by a PR creator, and maintains its value until reaching around 1200 PRs. The only PR creators who create more PRs are almost exclusively bots (e.g. greenkeeper bot), and they tend to create PRs that do not really reflect the experience gained by a human developer, and have a much lower chance of being accepted.  

\subsubsection{Dependence on previous social proximity , containing an issue fix, and presence of a dependency}
We do not have a variation of the PR acceptance probability per se for the three categorical variables, but the partial dependence plots show that having a previous PR accepted in the repository, a dependence between the package and the projects the PR creator contributed to, and an explicit mention of an issue fix tend to have a positive impact on the PR acceptance probability.

\hypobox{Answering RQ3: The variation of the PR acceptance probability and the measures we presented in this study are depicted using the partial dependence plots (Figure~\ref{t:pplot}) and the likely cause for such variation is discussed in detail. Overall, we observe that the nature of the variation is often nonlinear with peaks and troughs, something that can't be captured adequately by simple regression models, which was used by most of the previous studies to describe the nature of the relationships}


\vspace{-10pt}
\section{Limitations}\label{s:limit}
In our study, we focused only on the more popular NPM packages, which constitute less than 0.5\% of the entire NPM ecosystem. Although these are the packages that see the most amount of activity, and should, therefore, be of interest to most of the practitioners and researchers in the field, some of the factors found to significant for these packages might not be so important in the PR acceptance scenario for the less popular packages. 

As noted by one of the reviewers, sometimes a project might not use the conventional method of merging PRs, but implement some type of workaround per their convenience. Our study does not account for such situations, however, we didn't find any clear evidence of such activity for the projects we studied.

The result of this paper might not be applicable as-is to other software ecosystems, since every ecosystem has their norms and characteristics which is impossible to account for when looking into only one ecosystem. Future studies are needed to determine the generality of our findings.

Some of the characteristics (e.g. see Section~\ref{ss:partial}) observed in our study could be due to the presence of bots in the dataset that behave differently from human developers~\cite{dey2020detecting,dey2020exploratory}. Being able to get rid of them (e.g. by using the bot detection method proposed by~\cite{dey2020detecting}) would further improve the accuracy of our findings, which we plan to do as a future work. Disambiguation of the PR creator's IDs (e.g. by using the method proposed by~\cite{fry2020dataset}) can also help improve the accuracy of the result.
Another important topic that could be addressed by future work is finding out if the PR integrators actually find these signals to be useful and identify any factors we might have missed here.

\vspace{-10pt}
\section{Conclusion}\label{s:conclusion}
In this study, we aimed to examine the effects of various social and technical factors on the quality of a PR (its probability of being accepted). We formed seven hypotheses that replicate findings in prior work and also pose additional propositions that reflect the ecosystem-wide 
concerns. We fit logistic regression models that show statistically significant relationship of PR acceptance and 15 hypothesised predictors. We also predict the acceptance of future PRs with AUC of 0.94. Finally we explore the functional relationship between the predictors and the probability of PR acceptance and find it to be nonlinear and even non-monotone in many cases. Our key findings are listed in the answers to the research questions presented in the study.

Our findings have both theoretical and practical implications. The
accuracy of our PR acceptance model increases the likelihood of
successful practical applications that range from tools that
support PR integrators to tools that help the PR creators to
tailor their contributions to the form resembling that of the PRs
that are most likely to be accepted by a specific project. As the NPM
ecosystem and other OSS ecosystems depend on contributors to
maintain growth and code quality, we hope that the results of our
work would help these ecosystems to sustain evolution and the high
quality of the code.

\vspace{-10pt}
\section*{Acknowledgement}
The work has been partially supported by the following NSF awards:
CNS-1925615, IIS-1633437, and IIS-1901102.

\balance
\bibliographystyle{ACM-Reference-Format}
\bibliography{sigproc} 

\end{document}